\documentclass[twocolumn,showpacs,aps,prb,showpacs,altaffillsymbol,groupaddress]{revtex4}

\usepackage{graphicx}
\usepackage{dcolumn}

\begin{document}


\title{Intrinsic point defects and volume swelling in ZrSiO$_4$ under irradiation}

\author{J.M. Pruneda}
\email[Corresponding Author:]{mpru02@esc.cam.ac.uk}
\author{T. D. Archer}
\altaffiliation{Present address: Physics Department, Trinity College, Dublin, Ireland.}
\author{Emilio Artacho}
\affiliation{Department of Earth Sciences, University of Cambridge
Downing Street, Cambridge, CB2 3EQ, United Kingdom}





\date{\today}

\begin{abstract}
The effects of high concentration of point defects in crystalline ZrSiO$_4$ 
as originated by exposure to radiation, have been simulated using first principles 
density functional calculations.  Structural relaxation and vibrational 
studies were performed for a catalogue of intrinsic point defects, with 
different charge states and concentrations.  The experimental evidence 
of a large anisotropic volume swelling in natural and artificially 
irradiated samples is used to select the subset of defects 
that give similar lattice swelling for the concentrations studied,  
namely interstitials of O and Si, and the anti-site Zr$_{\text{Si}}$, 
Calculated vibrational spectra for the interstitials show additional 
evidence for the presence of high concentrations of some of these defects in 
irradiated zircon.

\end{abstract}

%

\maketitle


\section{Introduction}

Large concentration of defects in materials produce strains that 
affect the crystalline lattice structure, which can  undergo 
amorphization, change into a different crystalline phase, or remain in 
the same crystal phase with changes in the lattice parameters.  In 
addition to these changes, the electronic properties are affected in a
way that can produce substantial differences in the physical and chemical 
properties of the original material.
Radiation effects in ceramic materials have been subject of a 
considerable number of studies due to their use in nuclear
fusion applications, and in the stabilization and immobilization 
of nuclear high-level waste\cite{Weber98,Grimes}.  Defect accumulation and its
consequences for the degradation of the mechanical properties 
have to be understood in order to control possible leaching, creep and 
fatigue of these materials.  
A number of issues have to be investigated, 
such as the stabilities of possible defects, their structure,
migration energies for interstitials and vacancies, 
displacement threshold energies, defect interactions and clustering, 
and their electronic structure.

Zircon (ZrSiO$_4$) has attracted considerable attention due to several
physical properties of interest, such as high permitivity (zirconium 
silicates are candidates as a gate dielectric in future silicon 
field effect transistors\cite{Wilk}), low thermal conductivity and 
high chemical stability.  It has also been proposed as a ceramic host 
for the immobilization of nuclear waste due to its high ability to 
accommodate actinides by substitution of Zr, and its high resistance 
to corrosion\cite{Weber98,Ewing95}.  Natural zircon may contain 
uranium in concentrations of up to 20,000 ppm, and samples with up to
$~$4 billion years have been dated.  Thus, even if possibly not the 
best final material for nuclear waste encapsulation, it is an
ideal system to study the effects of radiation damage.

After a radioactive impurity undergoes an $\alpha$-decay transition, 
strong damage is produced in the crystalline structure.  The heavy 
recoiling atom (tens to hundreds of keV) propagates through the material 
losing energy mainly through atomic collisions, thus producing large 
cascade collisions and in consequence, amorphized regions.  The highly 
energetic $\alpha$ particle ($\sim$5 MeV) that propagates in the opposite 
direction to the recoil atom, dissipates most if its energy 
by ionization processes, and a relatively smaller number of atomic 
displacements mainly at the end of its path.

When the dose of $\alpha$-decays is not too large there is a coexistence 
between the amorphized (metamict) region and the crystalline phase, 
the latter containing point defects.  Barriers to recombination cause the
accumulation of large concentration of these defects, resulting
in changes of the unit cell volume and shape.  This swelling is measured from
the shifts in the reflection angle in X-ray diffraction (XRD) and it is
clearly distinct from the one produced by the amorphous region, observed 
by changes in the total density.

Experiments based on natural 
samples, Pu-doped zircon and ion-beam irradiation, show a unit cell 
volume expansion of up to 5\%.  This volume swelling is anisotropic, 
giving an increase of $\sim$1.5\% along the $ab$-axis, and $\sim$2\% 
along the $c$-axis\cite{H&G}.  Over geological periods the recovery of 
a large fraction of point defects is observed in natural zircons 
(as opposed to Pu-doped samples\cite{H&G,Weber93}), with preferential 
self-annealing of the $ab$ plane\cite{Rios}.

Characterization of structural changes in the crystalline 
phase\cite{Rios} show that the dodecadeltahedron ZrO$_8$ is 
the less stable unit in zircon, and the defects are expected 
to be more closely related to it, as the Si atoms are well 
confined in the SiO$_4$ tetrahedra.  Theoretical studies of 
neutral point defects show that the oxygen interstitial is the 
most stable one\cite{Crocombette}. A non negligible concentration 
is expected at thermal equilibrium. 
The energies required to form native defects in zircon were 
investigated by molecular dynamics simulations of primary-knock-on-atom 
collisions with empirical potentials\cite{PKA}.  
It was shown that the effective energy required to displace O from 
its crystal position is $\sim$17 eV, whereas it is larger for Si 
and Zr ($\sim$48, and $\sim$60 respectively).

The previous theoretical studies of point defects in zircon were focused
on the energetics of neutral defects.  However, due to ionization processes 
produced by the $\alpha$-particle, and the fast that the resulting defect 
landscape affects the electronic chemical potential, the defects can be charged.  
The accumulation of charged defects produce electric fields 
that would change the kinetics of accumulation and diffusivity.
In the present work, we use first principles electronic structure 
calculations to study the influence of very high concentration of neutral and
charged point defects on the crystalline structure of ZrSiO$_4$.  
We perform structural relaxations, studying electronic properties,
as well as some spectroscopic calculations of a wide variety of defects, forming
a quite complete catalogue that is compared with experimental data.  
The effects of order/disorder of these defects have not been considered in the
present study, where periodic supercells were used to describe them.


\section{Methodology}
Calculations of the electronic properties of ZrSiO$_4$ are performed with
the self-consistent {\it ab initio} {\sc siesta} method\cite{siesta}, using
Density Functional Theory (DFT)\cite{DFT} within the Local Density 
Approximation (LDA)\cite{ca}.
The core electrons are replaced by norm-conserving 
pseudopotentials\cite{tm2} in the Kleinman-Bylander form\cite{kb}, 
generated with Zr(4s$^2$,4p$^6$,4d$^2$,5s$^2$), Si(3s$^2$,3p$^2$), 
and O(2s$^2$,2p$^4$) atomic valence configurations.
The 4p electrons in Zr were explicitly included in the calculations as 
semicore states, due to the large overlap with the valence states.
The wave functions are described with linear 
combinations of strictly localized numerical atomic orbitals\cite{basis}.  
We used a single-$\zeta$ basis set for the semicore states of Zr, 
and double-$\zeta$ plus polarization for the valence states\cite{optZr}.  The 
charge density is projected on a real space uniform grid with an 
equivalent plane-wave cutoff of 227 Ry, to calculate the 
exchange-correlation and Hartree matrix elements.

To simulate the defect structures, the host crystal is represented 
by a supercell generated by repetition of the conventional tetragonal 
unit cell (4 formula units).  The point defect is then introduced 
inside this supercell (adding atoms for interstitial, removing 
atoms for vacancies, etc), that have to be of the right size to 
describe the sought large defect concentrations.  
We use host supercells with 24, 48, and 96 atoms.  The supercell 
with 48 atoms is a repetition of the tetragonal cell along the $c$ axis, 
and the one with 96, is a repetition along the $a$ and $b$ axis.
In this way, different concentrations of defects can be simulated by 
changing the number of repetitions in the cell.  The supercells with 
24 and 48 atoms were already used by Crocombette in his study of points 
defects, and he estimated that unreal physical interaction between 
periodic images of the defects was of the order of 0.5 eV.  For the high
concentrations we are interested in, these interactions between defects 
are expected and only their periodic arrangement might be unrealistic.
The use of more than one defect per supercell can alleviate this 
approximation, and will be discussed later.

Simulating charged defects with periodic boundary conditions, 
requires handling with care the long-range coulombic interactions.
The unphysical divergence in the energy coming from the long range 
Coulomb interactions of a periodic array of charges 
is compensated by a uniform electron-charge neutralizing 
background\cite{payne}.
The minimum-energy structure for each defect was obtained by relaxing 
with a conjugate-gradient minimization for the forces and stresses.  
The lattice vectors and the atomic positions were allowed to relax 
until atomic forces and stresses from the electronic structure calculations
were smaller than 30 meV/\AA\ and 6 meV/\AA$^3$, respectively.
The formation energies of the defects in the different charge states 
are function both of the electron chemical potential $\mu_e$, and of the 
atomic chemical potentials for the species involved in the defect, and
will be discussed elsewhere\cite{Edefects}.

The local vibrational modes (LVM) associated to the defects are 
obtained from the dynamical matrix
in the relaxed structure, by finite differences, i.e. computing 
the energies with each atom displaced by $\sim\pm$0.016\AA\ along 
each of the cartesian coordinates.  The eigenvalues and eigenvectors 
of the dynamical matrix give the vibrational 
frequencies and eigenmodes in the harmonic approximation.  Previous 
studies show a deviation of $\sim$5\% for vibrational frequencies of 
point defects in semiconductors.

The Born-effective charge tensors, $Z_{ij,\alpha}^*$, 
are used to determine the infrared intensities, and the splitting 
between longitudinal (LO) and transverse (TO) optic modes\cite{LOTO}.  
They are computed by finite differences as the change in the $i$ 
component of the polarization (computed with the Berry phase 
formalism\cite{Vanderbilt}) when the atom $\alpha$ is displaced 
along the $j$ direction.  


\section{Crystal $ZrSiO_4$}

\begin{figure}[t]
\includegraphics[scale=0.75]{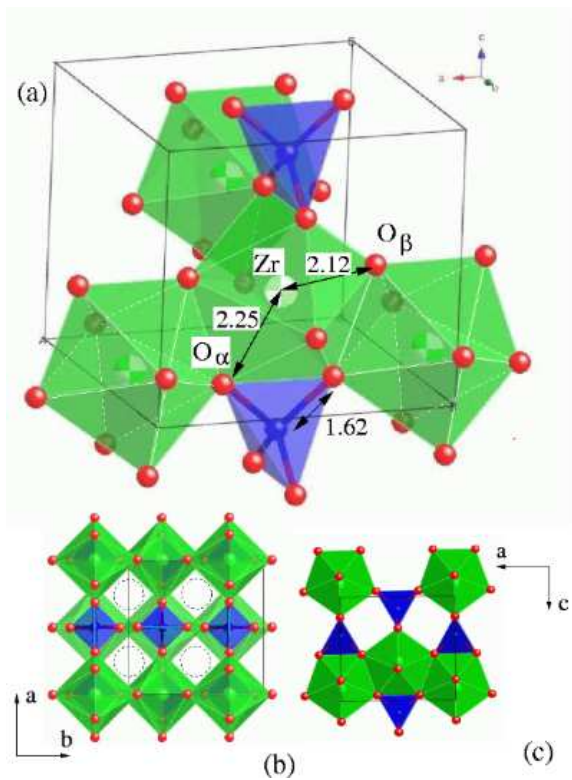}
\caption{({\it Color Online})
(a) Edge-sharing Si-tetrahedra and Zr-dodecadeltahedra
along the $c$-axis, corner sharing in the $ab$-plane. The 
dodecadeltahedra also have edge-sharing with each other along 
the $ab$-plane. (b) and (c) show the projection of the unit 
cell in the $ab$ plane, and in a plane perpendicular to the $b$ axis 
(the second layer of polyhedra has been removed in (c) for a better 
visualization). The dashed circles in (b) show the empty spaces in which we
place the interstitials.}
\label{polyhedra}
\end{figure}

Zircon (ZrSiO$_4$) has tetragonal $I4_1/amd$ space group. 
The structure consists of alternating SiO$_4$ tetrahedra 
and ZrO$_8$ triangular dodecadeltahedra sharing edges and 
forming chains parallel to the crystallographic $c$ axis 
(Fig. \ref{polyhedra}).  A body-centred unit cell can be chosen, 
containing four formula units.  The structure is fully described with 4 
parameters (see table \ref{lattice}): The lattice constants $a$ 
and $c$, and the internal parameters $y$ and $z$. Zirconium and 
silicon atoms are located in the $4a$ and $4b$ Wyckoff positions, 
$(0,\frac{3}{4},\frac{1}{8})$ and $(0,\frac{1}{4},\frac{3}{8})$,
respectively, and the oxygen atoms in the $16h$ sites (0,y,z).
Each O is shared between one Si and two Zr atoms, at distances 1.62, 
2.12 and 2.25\AA\ respectively.  These 4 atoms are in the O-site's 
mirror plane that can be perpendicular to the $a$ or $b$ axis.
The relaxed parameters calculated are in good agreement with 
experimental values.

\begin{table}[b]
\caption{Calculated structural parameters for crystalline ZrSiO$_4$ 
compared to experimental values. }
\begin{tabular*}{8cm}{c@{\extracolsep{\fill}}ccc}
       &  This work    &   Ref. [\onlinecite{Gonze}] 
                                           & Experiment\cite{latt} \\
\hline
Volume &  129 (\AA$^3$)&   127		   &  131       \\
 a     &  6.59 (\AA)   &   6.54		   &  6.61      \\
 c     &  5.96 (\AA)   &   5.92		   &  5.98      \\
 y     & 0.068         &   0.0645	   &  0.0661    \\
 z     & 0.184         &   0.1945	   &  0.1953    \\
\\
d(Si-O)     & 1.62    &   1.61             &  1.62  \\
d(Zr-O$^{\alpha}$) & 2.25    &   2.24             &  2.27  \\
d(Zr-O$^{\beta}$)  & 2.12    &   2.10             &  2.13  \\
$\widehat{O\textendash Si\textendash O}$ & 96$^\circ$ & 97$^\circ$ & 97$^\circ$\\
 & 116$^\circ$ & 116$^\circ$ & 116$^\circ$\\
\hline
\end{tabular*}
\label{lattice}
\end{table}


The Born effective-charge tensor is shown in table \ref{BC} and is 
in excellent agreement with the results obtained with linear response 
calculations\cite{Gonze}.   Because of the local symmetry in the Zr 
and Si sites, 
the Born charge tensors for these atoms are diagonal, with only two 
independent components, Z$^*_\parallel$ and Z$^*_\perp$ parallel 
and perpendicular to the crystallographic $c$-axis, respectively. 
The large value of Z$^*_{Zr}$ and Z$^*_{Si}$ with respect to the 
nominal ionic electronic charge is interpreted in terms of mixed 
covalent-ionic bonding due to hybridization with oxygen\cite{Gonze}.  
For O, the Born charge tensor is anisotropic, with the charge in the 
mirror plane being larger than the nominal ionic charge (Z=-2), and 
the tensor component perpendicular to the mirror plane being smaller.

\begin{table}[b]
\caption{Born effective charge tensors for Zr, Si, and O atoms. 
The Born charge for O, correspond to the atom located at $(0,u,v)$, with
the mirror plane perpendicular to $x$. 
For Zr and Si, only the diagonal elements are shown.  For O, the whole
charge tensor is presented, as well as the corresponding eigenvalues.
See also Ref. [\onlinecite{Gonze}]}
\begin{tabular*}{8cm}{c@{\extracolsep{\fill}}cc}
\hline
\hline
\\
Z$^*_{Zr}$ &  $\begin{pmatrix}{5.33 & 5.33 & 4.66} \end{pmatrix}$ & \\
Z$^*_{Si}$ &  $\begin{pmatrix}{3.21 & 3.21 & 4.38} \end{pmatrix}$ & \\
Z$^*_{O}$  &  $\left(\begin{matrix}{-1.19 &   0   &   0   \\
                                 0  & -3.11 & -0.15 \\
                                 0  & -0.34 & -2.26 } \end{matrix}\right)$ &
$\left[\begin{matrix}{	-1.19 \\
			-3.16 \\
			-2.20 } \end{matrix}\right]$ \\
\\
\hline
\end{tabular*}
\label{BC}
\end{table}

Seven infrared active modes (3A$_{2u}$+4E$_u$) are predicted by 
group theory in ZrSiO$_4$ at $\Gamma$.  The induced polarization 
for E$_u$ modes is perpendicular to the $c$ axis, while A$_{2u}$ 
modes are observed for electric field vectors parallel to c.
The spectrum between 650 and 1400 cm$^{-1}$ is characterized 
by Si-O stretching vibrations (E$_u$(4) and A$_{2u}$(3) modes 
respectively \cite{Zhang2001}) and the O-Si-O bending 
modes E$_u$(3) and A$_{2u}$(2).   
Our computed frequencies for these modes are 851, 969, 414, and 
602 cm$^{-1}$ respectively, with a maximum deviation of 3.8\% 
with experiments.  The modes below 400 cm$^{-1}$ are 
related to SiO$_4$ group motions against Zr atoms, and motions 
of Zr atoms.

\begin{table}[b]
\caption{Frequencies for IR active modes in ZrSiO$_4$ (in cm$^{-1}$).}
\begin{tabular*}{8cm}{c@{\extracolsep{\fill}}ccc}
\hline
 Mode  & This work & Ref. [\onlinecite{Gonze}] & Expt.\cite{Zhang2001} \\
\hline
\hline
A$_{2u}$(1) & 371 & 348 & 338 \\
A$_{2u}$(2) & 602 & 601 & 608 \\
A$_{2u}$(3) & 969 & 979 & 989 \\
E$_{u}$(1)  & 295 & 285 & 287 \\
E$_{u}$(2)  & 402 & 383 & 389 \\
E$_{u}$(3)  & 414 & 422 & 430 \\
E$_{u}$(4)  & 851 & 867 & 885 \\
\hline
\label{IR-modes}
\end{tabular*}
\end{table}

\section{Point defects}

We have considered a catalogue of neutral and charged intrinsic 
defects, that includes interstitials (X$_i$) and vacancies (V$_X$) of the three 
elements ($X$) present in zircon, Zr and Si anti-site defects (Zr$_{\text Si}$, 
and Si$_{\text Zr}$), Frenkel pairs ($X_{FP}$), and other combinations of 
interstitials and vacancies.

\subsection{Volume swelling}

\begin{figure*}[t]
\includegraphics[scale=0.70]{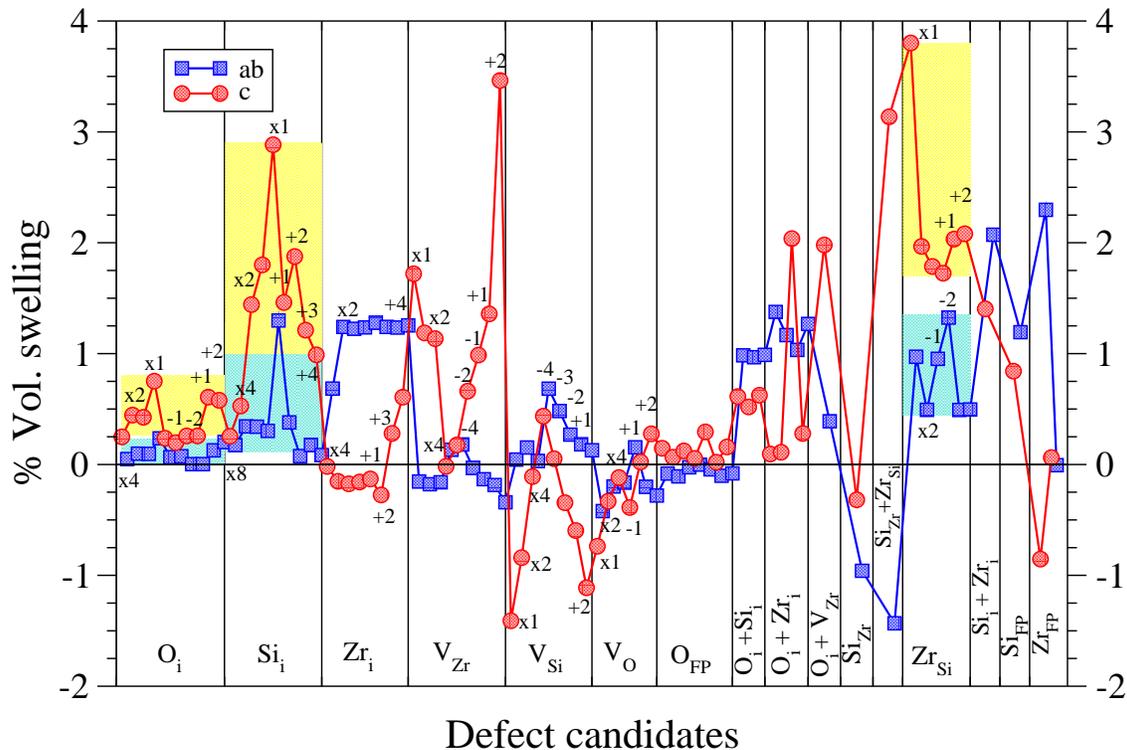}
\caption{({\it Color online}) Lattice parameter swelling (in \%) for 
the set of defects 
considered in this work.  An average in the basal (ab) plane is done 
to compare with the relaxed crystalline structure.  X$_i$, V$_X$,X$_{FP}$ 
denote interstitials, vacancies and Frenkel pairs of X, respectively. 
The different charge states are denoted by $\pm q$
and the different concentrations of defects by $\times n$, with  
$n$ denoting the number of repetitions of the tetragonal unit cell 
(cells with 24, 48 and 96 atoms, for $n=1,2,4$ respectively). Unless 
stated, the neutral defect with $n=2$ cell is considered.}
\label{swelling}
\end{figure*}

\begin{figure}[t]
\includegraphics[scale=0.45]{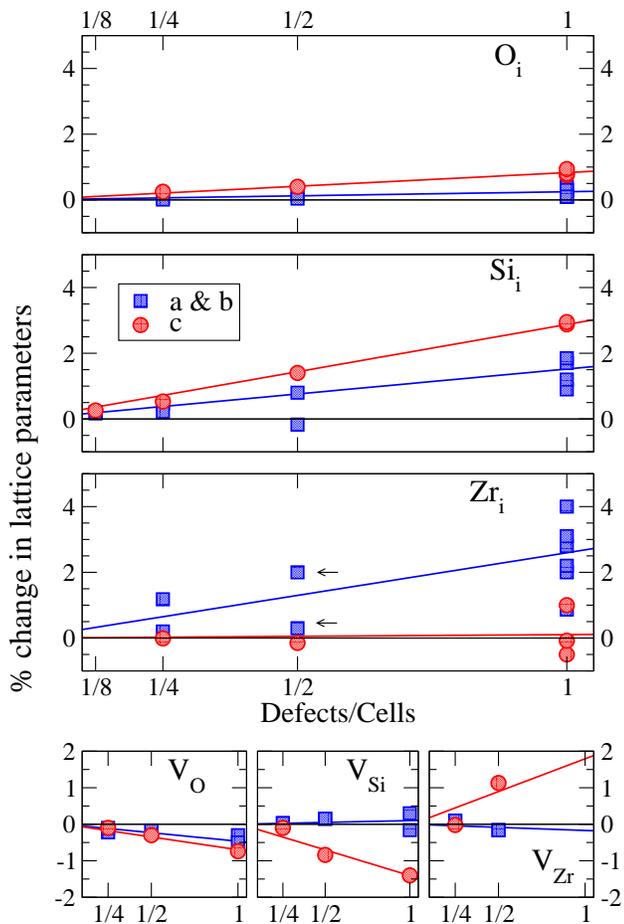}
\caption{({\it Color online}) Change in lattice constants (\%) as a function of the 
concentration of defects, for interstitials and vacancies of the three 
constituent elements. Results are shown for 1 defect for each 
$N$-repetitions of the tetragonal unit cell, with $N=1,2,4$ and 8.
The small arrows show the anisotropic swelling in the $ab$ plane 
for Zr$_i$. The different points for 1 defect/cell refer to different 
disorder realizations.}
\label{concentration}
\end{figure}

The effect of point defects in the lattice structure after relaxation 
is summarized in Fig. \ref{swelling}, where we show the change in the 
lattice parameters relative to the neutral structure for the defects studied.  
Some defects produce a considerable distortion of the crystalline lattice, 
whereas others such as oxygen interstitial, or the vacancies have a much 
smaller effect.  The magnitude of the distortion depends on the 
concentration of defects (supercell size in our study). For 
the values of interest in this work, a clear trend 
is observed, with an almost linear dependence between the distortion 
and the concentration (see Fig. \ref{concentration}).
For the highest concentration studied (one defect per unit cell), we
explored the effect of defect disorder by introducing two point defects
in the supercell made from two unit cells.  Fig. \ref{concentration} shows
a dispersion in the swelling due to the different configurations 
studied in this way. Even if the effect is sizeable it does not affect 
our arguments or conclusions, though further work in this direction 
will be needed.

In general, there is a considerable anisotropy in the relaxation, with 
different changes in the $c$-axis and in the $ab$-plane.
Experimentally\cite{H&G}, an increase in $c$ of $\sim$2\% is 
observed, while $a$ increases by a smaller amount ($\sim$1.5\%).  
Based in this evidence, we can focus 
our study on those defects that give a behaviour in the lattice
swelling similar to the one observed, eliminating those that 
produce a decrease in the cell parameters, or a larger change 
in $a$ than in $c$.  The defects with the stronger distortions 
are the interstitial of silicon, Si$_i$, and the anti-site of 
zirconium in the position of silicon, Zr$_{\text Si}$.
The calculations for charged systems were done with the supercell
of 2 tetragonal unit cells, repeated along the $c$-axis.  
The artificial coulomb interaction between cells should be then
more important in the $ab$-plane, but we observed that the lattice 
parameter $c$ is more strongly affected by the charge state than $ab$.

\subsection{Structural relaxation and electronic structure.}

\subsubsection{Interstitials}

Initially, we place the interstitial atoms in regions with empty space 
(see dashed circles in Fig. \ref{polyhedra}), and perform 
conjugate gradient minimization
of the forces and stresses.  We have explored different starting points,
showing here the relaxed structures with minimum energy.

\begin{figure}[b]
\includegraphics[scale=0.28]{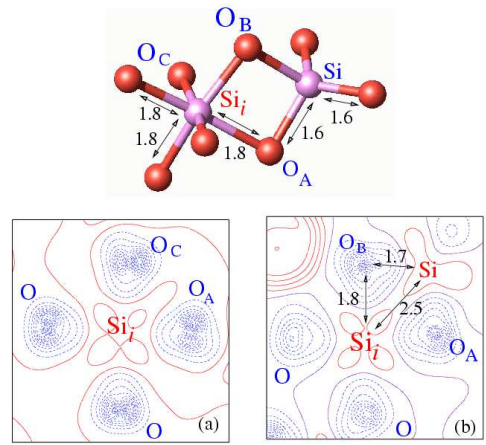}
\caption{({\it Color online}) Atomic configurations for silicon interstitial in the +4 charged state
and corresponding charge density-deformation 
contour plot in the plane defined by Si$_i$, O$_A$ and O$_C$ (a), 
and Si$_i$, O$_A$ and O$_B$ (b). 
Distances given in \AA\ and the contour plots are drawn
between -0.45 and 0.45$e$, with a step of 0.015$e$. Solid (dashed) lines 
correspond to negative (positive) values for the charge.}
\label{iSi}
\end{figure}

For Zr$_i$ and Si$_i$ the interstitial remains along the $c$-axis 
empty spaces (Fig. 1b).  Nevertheless, the lattice 
is strongly distorted and the SiO$_4$ and ZrO$_8$ polyhedra change 
to accommodate to the presence of the interstitial.
There is an anisotropic change in the lattice parameters.  While 
Zr$_i$ produces large orthorhombic distortion in the basal plane 
(with an increase of up to 3\% in the $a$ lattice parameter, 
for a concentration of 1 defect in two unit cells, and smaller 
distortions in the other parameters), the Si$_i$ distorts the
crystal mainly in the $c$-axis, but also in the other directions.
Upon removal of electrons from Si$_i$, new atomic structures are obtained,
with the interstitial causing oxygen atoms to approach, and neutralize the
charge around it.  In this way, the coordination of Si$_i^{+n}$ increases
with $n$ and for Si$_i^{+4}$ a six-fold coordinated structure results, 
with Si-O bonds slightly longer ($\sim$1.8\AA) than in the initial 
tetrahedral geometry.  In Fig. \ref{iSi} we show the electronic deformation
density (difference between the solid and atomic electronic clouds),
$\delta\rho$, around Si$_i$ in two different planes.  The geometries 
for positively charged Zr-interstitials are basically unchanged with 
respect to the neutral structure.

\begin{figure*}[t]
\includegraphics[scale=0.58]{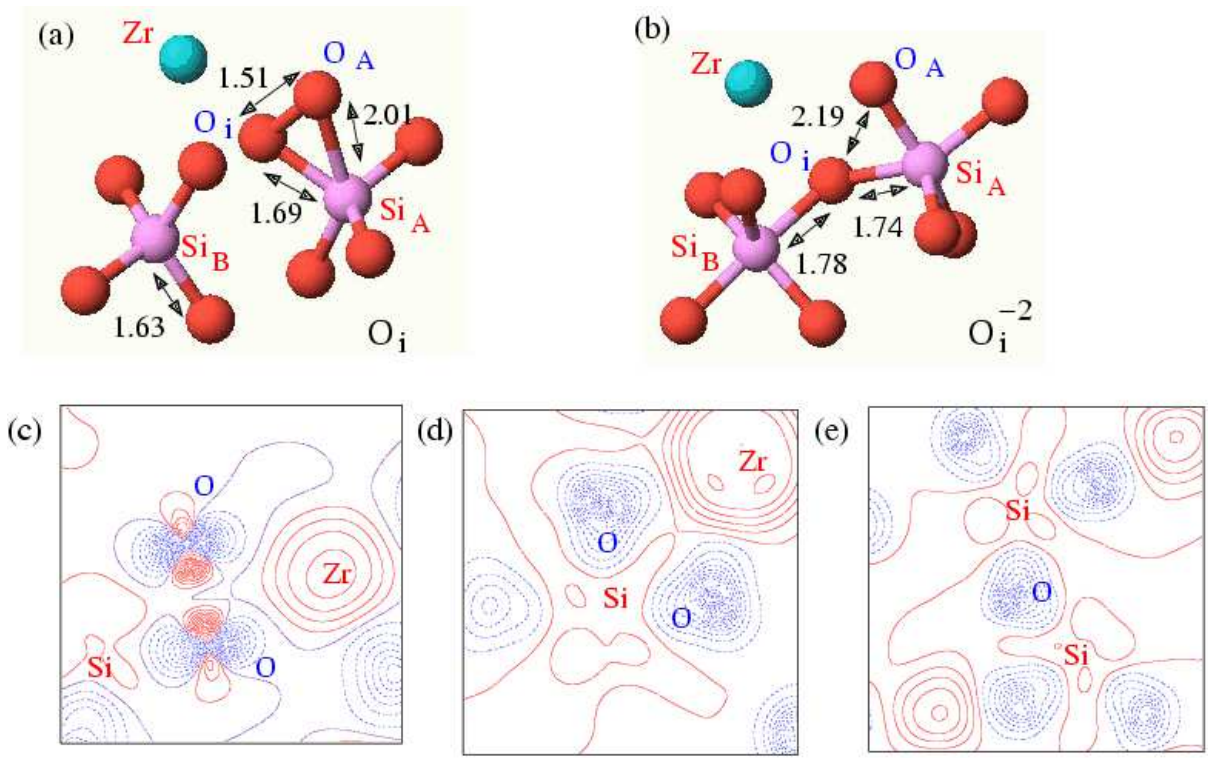}
\caption{({\it Color online}) Atomic configurations for oxygen 
interstitials in the neutral (a) and doubly negatively charged (b) 
states, and corresponding deformation density
contour plots in the plane defined by Si$_A$, O$_A$ and O$_i$ (c and d 
for neutral and charged configurations, respectively). 
(e) shows the contour plot in the plane through Si$_A$, O$_i$ and Si$_B$,
 in O$_i^{-2}$.  Distances given in \AA\ and the contour plots are drawn
between -0.45 and 0.45$e$, with a step of 0.015$e$. Solid (dashed) lines 
correspond to negative (positive) values of the deformation density.}
\label{i-O}
\end{figure*}

In agreement with previous 
theoretical studies\cite{Crocombette}, we observe that the neutral 
O$_i$ forms a ``dumbbell''-like structure (see Fig. \ref{i-O}a), 
with the axis almost 
perpendicular to the mirror plane of the oxygen in the crystal 
position. The distances 
between both oxygen atoms, and the neighbouring cations are shown 
in table \ref{dumbbell}.  A similar structure was also observed in 
calculations of interstitial O in ZrO$_2$\cite{Nieminen}.  The high 
ionicity of the crystal forces the interstitial to a position where 
the electronic density screens the coulomb interactions.  The covalent
bond to the interstitial oxygen reduces the strength of the Si-O bond, 
and the SiO$_4$ tetrahedra loses part of its charge.  Mulliken population
analysis shows that the charges in the interstitial and in the lattice oxygen
are similar, and slightly smaller (6\%) than in other oxygen atoms.

\begin{table}[b]
\caption{Interatomic distances in the dumbbell structure (in \AA), for
different concentration of defects (supercells with 24, 48 and 96 atoms). 
Charged defect is within the supercell containing 48 atoms.}
\begin{tabular*}{8cm}{c@{\extracolsep{\fill}}ccccc}
             & 24 & 48 & 96 & Ref.[\onlinecite{Crocombette}] &   O$_i^{-}$  \\
\hline
\hline
O$_i$-O$_A$  & 1.52  & 1.52  & 1.52  & 1.62   &  1.99 \\
O$_i$-Si     & 1.96  & 1.69  & 1.70  & 1.83   &  1.69 \\
O$_i$-Zr$_1$ & 2.29  & 2.24  & 2.23  & 2.24   &  2.15 \\
O$_i$-Zr$_2$ & 2.22  & 2.84  & 2.82  & 2.57   &  2.98 \\
O$_A$-Si     & 1.70  & 2.00  & 1.95  & 1.78   &  1.80 \\
O$_A$-Zr$_1$ & 2.24  & 2.23  & 2.30  & 2.20   &  2.32 \\
O$_A$-Zr$_2$ & 2.84  & 2.22  & 2.22  & 2.75   &  2.18 \\
\hline
\label{dumbbell}
\end{tabular*}
\end{table}

The structure of charged O$_i^{-}$ does not differ substantially 
from the neutral defect, with a slight increase of the O$_i^{-}$-O 
distance (see table \ref{dumbbell}).  On the contrary, the doubly 
charged interstitial displaces from the dumbbell structure, and 
forms a bridge between neighbour silicon atoms, both having now 
coordination five.  In this structure, the relaxed O$_i^{-2}$-O$_A$
distance (2.19 \AA) does not admit a bond between the oxygens, 
while Si$_A-$O$_i^{-2}$ and Si$_B-$O$_i^{-2}$ do (1.74 and 1.78 
\AA, respectively).  This can be seen in Fig. \ref{i-O}d, where 
$\delta\rho$ around the interstitial oxygen differs from
the one in the neutral defect (Fig. \ref{i-O}c).  The electronic cloud
along the Si$_A$-O$_i$-Si$_B$ directions (Fig. \ref{i-O}e) resembles 
the Si-O bond in the crystalline configuration.
The relaxation energy from the initial dumbbell configuration is 
of about 1.6 eV, while the energy of the neutral defect in the atomic 
structure of O$_i^{-2}$ is 2.9 eV higher.

\subsubsection{Vacancies}

In the V$_O$, the silicon atom moves slightly in the direction 
of the missing oxygen, and the remaining Si-O bond lengths are 
increased by 4\%.  A new electronic state is created in the gap,
localized around the vacancy.  In the charged defect
V$_O^+$, the Si atom moves back toward its initial position, and the
Si-O bonds have deviations smaller than 1\% with respect to the 
original Si-O tetrahedra.

The four oxygen atoms forming the tetrahedra around the silicon 
vacancy remain in their position, with only minor deviations 
towards their centre of mass ($\sim$-0.7\%).  For the 
V$_{\text Si}^{-n}$ configurations, the oxygen tetrahedron expands, with 
the distance from the centre of mass increasing by 2\% for $n=1$ and
by 5\% for $n=2$.  For the V$_{Zr}$ defect a similar effect is 
observed for the oxygen atoms forming the dodecadeltahedra.  
There is a considerable expansion of the V$_{Zr}-$O$^\alpha$ 
bond ($\sim$10\%), and a smaller change in the V$_{Zr}-$O$^\beta$
length ($\sim$+4\%), that explains the anisotropic change in 
lattice parameters.  The expansion of the V$_{Zr}-$O$^v$ bond increases for 
the negatively charged vacancies.  The V$_{Zr}-$O$^i$ length expansion 
decreases (increases) for negatively (positively) charged vacancies.

\subsubsection{Frenkel pairs and anti-sites}

Frenkel pairs of Zr and Si are not favourable because of the strong 
ionic interaction between the interstitial and the vacancy.  There is
a trend for the interstitial to move towards the open space of the 
vacancy, and a restitution was observed in the relaxation of 
several initial structures considered.  
In the case of O$_{FP}$, the
interstitial and the vacancy do not interact strongly, and the final 
configuration is similar to an isolated vacancy and a dumbbell interstitial, 
even when both are in the same Si-O tetrahedra.  

\begin{figure}[t]
\includegraphics[scale=0.38]{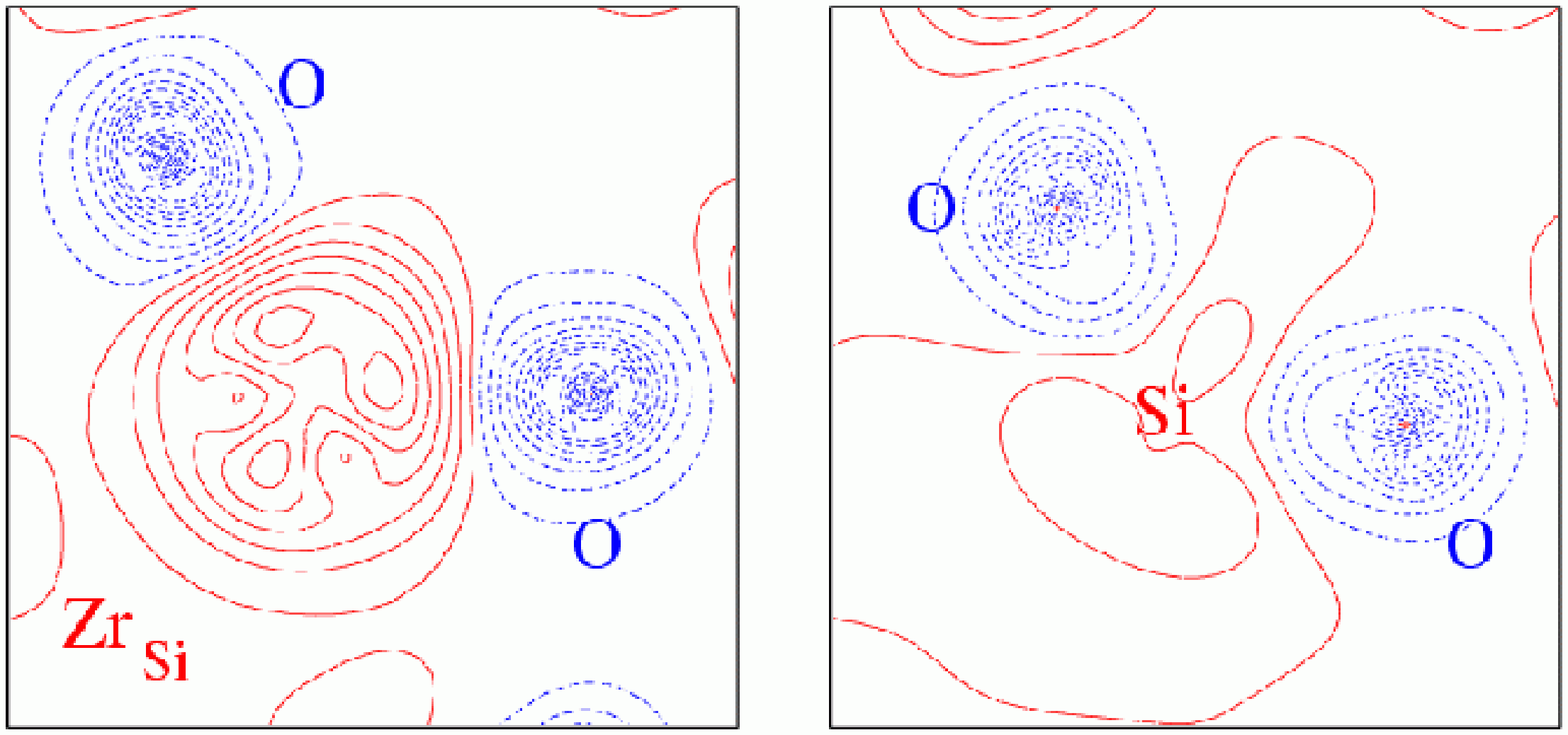}
\caption{({\it Color online}) Charge redistribution density around Zr$_{\text Si}$ and Si, in 
the tetrahedral plane.}
\label{Zr-O}
\end{figure}

In the Si$_{\text Zr}$ anti-site, the Si atom substitutes a Zr atom in the 
center of the dodecadeltahedron, with a reduction of the lengths 
(-4\% for Si-O$^i$ and -6\% for Si-O$^v$).  In Zr$_{\text Si}$, the
Zr forms a tetrahedron with four oxygen around it, with bond lengths 
of 1.9\AA (+17\% longer than the Si-O bonds).  The Si-O distances increase
slightly ($\sim$+1\%) close to the defect, due to the increase in the 
lattice parameters.  Fig. \ref{Zr-O} shows
the contour plot of $\delta\rho$ in the plane defined by Zr$_{\text Si}$ 
and a pair of oxygens in the tetrahedron.  In the charged configurations, 
the bond lengths do not change substantially.

\subsection{Vibrational properties}

The infrared (IR) reflection and absorption spectra of zircon is 
considerably affected by radiation damage. Changes in peak frequency, 
decrease in band intensity, and line broadening of the spectra of 
crystalline zircon can be related to changes in the lattice bond lengths. 
In addition, new features appear for highly damaged samples, 
with an increase in the IR intensity with increasing dose.  One of the 
new spectral features is the appearance of an absorption sharp peak near 
796 cm$^{-1}$ in moderately damaged zircon\cite{Zhang2000b}.  
This peak is in a gap
of the vibrational density of states (VDOS) of crystalline zircon.  Its 
origin is unclear, and it has been proposed that it could be due 
to a radiation-induced intermediate phase, or to distortions in the
boundaries between crystalline and amorphized regions\cite{Zhang2000b}.

\begin{figure}[b]
\includegraphics[scale=0.40]{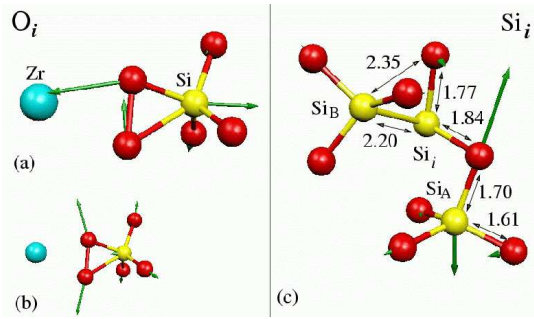}
\caption{({\it Color online}) Local Vibrational Modes for O$_i$ and Si$_i$ point defects. 
(a) and (b) show the eigenmodes for oxygen interstitial at 729 cm$^{-1}$ 
(IR active) and 870 cm$^{-1}$ respectively. (c) shows the IR active
mode for silicon interstitial at 734 cm$^{-1}$. The relatively 
high frequency of mode (b) is an indication of the bond strength 
between the oxygens in the dumbbell. (Distances in \AA)}
\label{LVM}
\end{figure}

We computed the vibrational properties for interstitials of O, Si and Zr, 
and for the anti-site defect Zr$_{\text Si}$, in the relaxed structures 
with 48 atoms.  
IR active local vibrational modes associated to the defects were observed 
for O$_i$ and Si$_i$ in a range of energies corresponding to the gap in 
the VDOS (729 and 743 cm$^{-1}$ respectively). The eigenmodes are shown in 
Fig. \ref{LVM}.  They mainly involve the stretching of Si-O bonds close 
to the interstitials. 
Other localized modes were observed for these point defects, but
they are not IR active.  

The crystal modes are affected by the presence of the defect.  
The changes are more important in Si$_i$, probably due to a 
stronger interaction with oxygen atoms around this interstitial. 
For O$_i$, the frequencies of the crystalline modes are almost 
unaffected (though some splittings are observed for the low frequency modes)
showing that the defective structure does not differ much from the 
crystalline one.  In the charged configuration (O$_i^{-2}$), the interstitial
is bonded to two silicon atoms, hence considerably affecting the vibrational
modes of the Si-O tetrahedra.  Some weak IR active modes appear in the 
gap for this defect.  Zirconium interstitial affects mainly the low 
energy part of the spectra (below 500 cm$^{-1}$), where the Zr-O 
interactions are more important.  The Si-O stretching modes are softer, 
probably due the increase in the distance between silicon tetrahedra, 
consequence of the large expansion in the $ab$-plane.  For the anti-site
defect Zr$_{\text Si}$, the high energy modes, involving the stretching 
of the tetrahedra bonds, are weaker.  The low energy modes,
mainly related to relative displacements of the Zr and Si-tetrahedra 
sublattices, are less affected.

\section{Discussion}
Experimentally, the change in the lattice parameters can be directly 
monitored using x-ray diffraction.  
Point defects would be randomly distributed through the crystal lattice.  
If the number of accumulated $\alpha$-decays is not too large 
($\sim2\times10^{18}\alpha/g$), and we assume that $\sim$220 atoms are 
displaced by each $\alpha$ particle\cite{Rios}, we can expect concentrations of 
$2\times10^{21}$defects/cm$^{3}$.  That is roughly one defect for each 
two unit cells, the kind of concentration considered in this work.
Similar concentrations have been used to study the effect of doping 
on the lattice parameters of semiconductors\cite{Si:As}, but the 
changes observed are considerably smaller ($\sim$0.19\%).  
We performed simulations of Si interstitials in the crystalline 
silicon structure with a supercell of 64 atoms (roughly the same 
concentration of defects), giving a volume swelling of 0.2\%, much 
smaller than the values obtained for zircon with the same approach.
In ceramic materials used for nuclear applications radiation 
swelling is more important ($\Delta V/V$ up to 8\% in SiC) and 
both defect accumulation and ratio between amorphized and crystalline 
phases play an important role\cite{SiC}.

When compared with experiments, our calculations at those 
concentrations, single out three candidates as possible originators of the 
lattice swelling: O$_i$, Si$_i$ and Zr$_{\text Si}$.  
Some concentration of these three defects should be there to account for the 
experimental findings, even if other defects can be present at lower 
concentrations.  In the following we discuss the compatibility of this 
conclusion with other facts known about the system.

For weakly damaged zircon, the fraction of crystalline phase is larger 
than the fraction of amorphized regions, and the diffraction maxima are 
weaker than in undamaged crystals.  The anisotropic thermal displacement 
factors have been used to quantify the effect of the radiation damage in 
zircon\cite{Rios}, indicating that the damage has more important 
effects on Zr and O atoms than on Si, which remain inside the SiO$_4$ 
tetrahedral units due to the stronger binding of Si with O.  
In contrast with these findings, calculations of threshold displacement 
energies\cite{PKA} (E$_d$) show that for neutral atoms
E$_d$(O)$\ll$E$_d$(Si)<E$_d$(Zr).  

However, it is important to emphasize that E$_d$ can change substantially 
with the charge state.  Charged defects can migrate through the lattice 
with a lower migration energy than neutral ones.  In fact, the atomic 
configurations for oxygen interstitials are similar to the
ones observed for ZrO$_2$ and HfO$_2$\cite{Nieminen,Hafnia}.  In these 
materials, oxygen is incorporated and diffuses in atomic form, acting as 
an electron trap, while neutral defects tend to form strong bonds with other 
lattice oxygens. The negative-U behaviour favors the appearance
of charged interstitials that are less tightly bound to other
oxygens.  

In a real material, we can expect all kind of defects being present, 
with the swelling produced by some of them being compensated with the 
deflation due to others.  The presence of interstitials
requires that vacancies are formed somewhere in the lattice, but the effect
of these vacancies on the lattice parameters is smaller (see Fig. \ref{swelling}).
The formation of Zr$_{\text Si}$ leaves a V$_{\text Zr}$ and a interstitial 
of Si.  The combination of these three defects is compatible with a 
larger swelling along the $c$-axis than in the $ab$-plane, as observed 
experimentally.
The formation energies for the point defects studied\cite{Edefects}, 
indicate that the most probable 
defects would be interstitials of oxygen and the anti-site 
Zr$_{\text Si}$.  Depending on the chemical potentials of each species 
($\mu_{\text O}$, $\mu_{\text Si}$ and $\mu_{\text Zr}$) 
interstitials of Si and Zr might be present.
The IR active mode at $\sim$730 cm$^{-1}$ would give further evidence for the
presence of oxygen, and maybe silicon, interstitials.  

$^{29}$Si nuclear magnetic resonance studies of damaged zircon indicate 
changes in the Si local environment towards a polimerization of the 
structure\cite{nmr} during amorphization.  This means that Si$-$O$-$Si 
bonds are formed in the amorphous region and the 
initially isolated SiO$_4$ tetrahedra are connected through an 
oxygen bridge.  This polymerization requires that on average the
number of oxygens per Si is smaller in the amorphized region, favouring the
presence of interstitial oxygens in the rest of the system.  
As we have seen O$_i^{-2}$ forms
bridging structures between Si atoms.  But also the presence charged
Si interstitials can give similar polymerization and change the chemical 
shifts of crystalline Si.

%

\section{Conclusions}
Simulations of high concentration of point defects in ZrSiO$_4$ have been
performed to study the effect of these on the lattice swelling under 
radiation damage.  Using different sizes for the host crystalline 
supercell, we obtain various concentrations of periodically repeated 
defects.  A roughly linear dependence between the swelling and the
defect concentration has been observed for these ordered defective 
crystal.  Based on experimental evidence of anisotropic swelling, 
we have selected a set of defects as good candidates to be responsible 
for the lattice expansion in
crystalline zircon.  Vibrational properties in radiation damaged samples are
also used to justify our conclusions.  These include interstitials of oxygen 
and silicon, and the anti-site defect Zr$_{\text Si}$.

\section{Acknowledgments}
This work was supported by British Nuclear Fuels (BNFL) and NERC.
We would like to thank E. K. H. Salje, M. T. Dove, K. Trachenko, M. Yang, S. Rios, 
I. Farnan for helpful discussions on radiation damage.

\end{document}